\documentclass[entropy,article,submit,moreauthors, showkeys, pdftex]{revtex4} 
 \usepackage{verbatim}
 \usepackage{mathtools}
 \usepackage{hyperref}
 \usepackage{amsmath}
 \usepackage{graphicx}
 \usepackage[utf8]{inputenc}
 \usepackage[english]{babel}
 \usepackage{xcolor}
 
\begin{document}

	\title{ Universality of   scaling  correlations   across probability distributions}
 	 \author{Vaibhav Wasnik}
 	 \affiliation{ 
 	   Indian Institute of Technology, Goa   \\
 	 	 }
 	 \email{wasnik@iitgoa.ac.in}
 
	\begin{abstract}
\section*{Abstract}  	
	 	Scale invariance and the resulting power law behaviours are  seen  in diverse systems. In this work we consider  translation, discretely rotational and scale invariant systems defined on a lattice,  such that the  variables defining the state at each lattice site take on the same range of finite values and are    picked   from   probability distributions that are otherwise   arbitrary. We show that the   exponent that describes the scaling of the two point correlation function in these systems   matches the scaling exponent of a  equilibrium statistical mechanical  model   at criticality.  This work therefore extends the concept of universality   in statistical mechanics  to   probability distributions that are not Boltzmannian. 
	\end{abstract}
		  

 \maketitle
 
	\section*{Introduction}
	Scale invariance  is seen in various  natural and artificial   systems \cite{krug}, in percolation \cite{percolation}, in vocalization sequences \cite{vocalization}, scale invariance in the repeating fast radio burst \cite{radiobust}, in examples of turbulence \cite{turbulence}, in river run offs \cite{river}, in	stock market crashes \cite{stock}, criticality in biological systems \cite{bialek} etc.

	  In  equilibrium statistical mechanics,  different systems in thermal equilibrium at the critical point become scale invariant and show similar behaviour. At thermal equilibrium,  the probability distribution of finding the system in a particular state is proportional to   $ e^{-H/k_B T}$, where $H$ is the Hamiltonian describing the system state. For  a system defined on a lattice $H$ could have the form 
	\begin{eqnarray}
	H = \sum_i J_i S_i + \sum_{ij} J_{ij} S_i S_j + \sum_{ijk} J_{ijk} S_i S_j S_k...
	\end{eqnarray}
	
	Here $S_i$  is a variable describing the state       at lattice site $i$.  At the critical point, the critical exponents of the system end up being  independent of   microscopic details  of the Hamiltonian describing the system and are instead given by parameters such as its symmetry, dimensions in which the system exists etc \cite{lubensky} . For example   value of $\eta$ in the correlation function 
	
	\begin{eqnarray}
	\langle S_I S_J\rangle_B -\langle S_I\rangle_B \langle S_J \rangle_B \sim  \frac{1}{\mid I-J\mid^{d+\eta - 2}},
	\label{scaling}
	\end{eqnarray}
is  independent of the form of the Hamiltonian, instead determined by things such as symmetry etc.  Here
\begin{eqnarray}
\langle  \phi \rangle_B  = \frac{ \sum_{\{ S_i \}} e^{- H/k_B T} \phi }{ \sum_{ \{ S_i\}  } e^{-H/k_B T } }.
\end{eqnarray}
In this write up, $ \mid I-J\mid$ denotes  the distance between lattice points labeled by $I$ and $J$.

In the case of continuous   equilibrium statistical mechanical systems  that are scale invariant we can define operators that scale as $\phi_i(\lambda x) = \lambda^\alpha \phi_i(x)$. If one improves this symmetry to conformal invariance (with restrictions on the form of $\lambda(x)$ depending on the dimensions of space),  the operators scale as  $\phi_i(\lambda (x) x) = \lambda(x)^\alpha \phi_i(x)$. 
 It is known that if the scaling dimensions of operators, their spin and three point correlations are know, then it is possible to evaluate all possible correlation functions in the theory, without knowing the Hamiltonian \cite{cardy},\cite{poland}.    Since the Hamiltonian  really defines the  the probability distribution as $e^{-H/k_B T}$,  the  natural question then appears   whether  even  the functional form of the probability distribution is of relevance, since the correlation functions can be evaluated without knowledge of the Hamiltonian. Another aspect to note is that    that   scale invariance of the two point correlation functions  is not  a property  only of equilibrium statistical models at criticality, but this scale invariance is seen in a multitude of  scale invariant systems   that are not in a thermal  equilibrium with a heat bath like non equilibrium lattice gases \cite{spohn}  \cite{gas}, non equilibrium quantum spin chains\cite{prosen}, retinal neurons also display critical behaviour \cite{bialek}  among other systems. Obviously these systems are not described by equilibrium statistical mechanics. The   question that naturally arises is whether we can make any statement about scaling correlations in scale invariant systems that do not depend on the specific functional form of the probability distributions defining these systems. If any such statement could be made, it would imply a universal feature about these scaling correlations that are seen in varied systems be them natural or artificially created.

In this work, we  consider translational, discretely rotational and scale invariant systems defined on a lattice, such that the variable that defines the state  at a  lattice site    takes on the same finite range of values. These variables are collectively     picked up from an arbitrary probability distribution. We show   that   the  scaling of the two point correlation function       is similar     to  the scaling   of  equilibrium statistical mechanical  models described by the Boltzmannian probability distribution  at the critical point.    This therefore extends the concept of universality beyond Boltzmannian probability distributions of statistical mechanics.  We would like to explicitly state that by a Boltzmannian probability distribution we mean a probability distribution that goes as $e^{H}$ where $H$ is a analytical function of the degrees of freedom in the problem. For example if we say that the variable $S_i$ represents the   state at lattice site $i$, $H$ will have a form like $H(\{ S_i \}) = a+ \sum_i a_i S_i + \sum_{ij} a_{ij} S_i S_j + \sum_{ijk} a_{ijk} S_i S_j S_k + ...$, where $a, a_i, a_{ij}, a_{ijk}...$ etc are constants. Hence an arbitrary probability distribution $P(\{S_i\})$ cannot be written as $e^{\ln P(\{S_i\}) }$ and then labelled as a Boltzmannian probability distribution. This is because $ \ln P(\{S_i\}) $ cannot generically equal $H(\{ S_i \}) = a+ \sum_i a_i S_i + \sum_{ij} a_{ij} S_i S_j + \sum_{ijk} a_{ijk} S_i S_j S_k + ...$, because $H(\{ S_i \}) = a+ \sum_i a_i S_i + \sum_{ij} a_{ij} S_i S_j + \sum_{ijk} a_{ijk} S_i S_j S_k + ... $'s can be analytically continued to the full complex plane, by continuing $S_i$'s to the whole complex plane,  with no branch cuts needed to define $H(\{ S_i \})$,  however  $ \ln P(\{S_i\}) $  continued to the entire complex plane will have to be defined using branch cuts. Hence a functional relationship $\ln P(\{S_i\}) = a+ \sum_i a_i S_i + \sum_{ij} a_{ij} S_i S_j + \sum_{ijk} a_{ijk} S_i S_j S_k + ... $ is not possible in general, despite being possible for isolated cases. However, in this work we will prove our assertion stated in the beginning of this paragraph that for any   translational, discretely rotational and scale invariant systems defined on a lattice, where the state at every lattice site takes on the same range of finite values and are picked up from an arbitrary probability distribution, the  scaling of the two point correlation function       is similar     to  the scaling   of  equilibrium statistical mechanical  models at criticality.

 We summarize the arguments of the proof for the interest of the reader. Proposition 1, starts with the attesting  that if all possible measured observables are translationally invariant, the probability distribution describing the system should be translationally invariant. Proposition 2, states that translation invariance and discrete rotational invariance automatically makes the two point correlation function fully rotational invariant and hence functionally dependent on distance between the two points on the lattice. Scale invariance then sets the    correlation function $C(|a-b|)= \frac{C}{|a-b|^\alpha}$. In proposition 3, we  claim that translation invariance of the probability distribution implies that the probability distribution can be written as a function of translation invariant terms, that cannot be expressed in terms of each other.  This information allows us to write the probability distribution as a Fourier transform as
 \begin{eqnarray}
 f(\{S_i\}) &=&   \int \Pi_i dJ_i  \quad a(J_1, J_2, J_3... )  \quad e^{i J_1  \sum_i S_i + i  \sum_{m,n }J_2^{m,n }  \sum_i S_{i+n} S_{i+m} + i\sum_{m,n,p}J_3^{m,n,p}  \sum_i S_{i+p} S_{i+m} S_{i+n} ....} + (*),\nonumber \\
 \end{eqnarray}
 
  We can hence write the two point correlation function   $ C(|a-b|) = \frac{C}{|a-b|^\alpha}$ for the scale invariant system in terms of  correlation functions $C(a,b)_{J_1,J_2...}$ evaluated using the probability weight 
  $ e^{i J_1  \sum_i S_i + i  \sum_{m,n }J_2^{m,n }  \sum_i S_{i+n} S_{i+m} + i\sum_{m,n,p}J_3^{m,n,p}  \sum_i S_{i+p} S_{i+m} S_{i+n} ....} $.
    Rotational invariance of $C(a,b)_{J_1,J_2...}= C(|a-b|)_{J_1,J_2,J_3..}$ is emphasised. Proposition 4, then considers the case when distance between points $a$ and $b$ is orders of magnitude larger than the lattice spacing to show that in that limit $	 C(|a-b|)_{J_1,J_2,J_3..} \sim \frac{e^{ - \lambda_ {J_1,J_2,J_3..}|a-b| }}{|a-b|^{\alpha_{J_1,J_2,J_3..} }}$. Proposition 6, still looks in the limit when the distance between points $a$ and $b$ is orders of magnitude larger than the lattice spacing to show that wherever $C(|a-b|)_{J_1,J_2,J_3..} $ makes a non-zero contribution to $C(|a-b|)$, $ \alpha_{ J_1, J_2, J_3..}$, has to equal $\alpha$.  Finally in Proposition 7, we show that atleast one of the $\lambda_ {J_1,J_2,J_3..}$ has to equal zero, implying that $\alpha$ also equals the  scaling of a correlation function of a system at criticality described by a Boltzmannian probability distribution, completing our proof. Some of the propositions stated may have been known earlier but we   offer a proof for them for the sake of completeness. However all the propositions   build up to prove the main assertion of the paper which is a new addition to literature, that the scaling coefficient of the two point function for a translational, discretely rotational and scale invariant system on a lattice described by an  arbitrary probability distribution is similar to that of a translational, rotational and scale invariant statistical mechanical system at equilibrium.

\section*{Conventions}
Consider a system defined on a lattice. Let us label   lattice sites by vectorial indices $i$. Let $S_i$ be the  variable defining the state  at lattice site $i$. 
If the lattice is one dimensional $i$ is an integer.  If the lattice is two dimensional $i = (x_i,y_i)$ where $x_i,y_i$ are integers. If the lattice is three dimensional $i = (x_i,y_i,z_i)$ where $ x_i,y_i,z_i$ are integers. The lattice in question is constructed by primitive translation vectors and hence possesses a discrete rotational symmetry. 


We also note that $i$ appearing as a part of a subscript is a label and  not to be confused with $i = \sqrt{-1}$  that appears in exponentials of a Fourier transform in this paper.

Define translation as the following operation: Consider any variable $\{Z_i\}$ defined on the lattice.  Assign the value of the variable for lattice site $i+1$ the value of the variable of lattice site $i$, with $i$ running over all lattice sites. Denote this operation by  $Z_i \rightarrow Z_{i+1}$.  A quantity being translation invariant then would imply that the quantity is invariant under translations.

In the above when we say lattice site $i+1$ we are implying  the number $i+1$ if the system is one dimensional,  lattice site  $(x_i+1,y_i)$  along with $(x_i,y_i+1)$ if $i = (x_i,y_i)$ for a two dimensional lattice or implying the lattice site $ (x_i+1,y_i,z_i)$ along with  lattice sites $ (x_i,y_i+1,z_i)$ and $ (x_i,y_i,z_i+1)$, if $i = (x_i,y_i,z_i)$ for a three dimensional lattice.  $(1,0), (0,1)$ and $(1,0,0), (0,1,0), (0,0,1)$ are the primitive translation vectors of the lattice in two and three dimensions respectively. Nowhere it should be assumed that the primitive translation vectors have to be perpendicular to each other in space.

To elaborate, let us consider the system in two dimensions. The operation  $Z_i \rightarrow Z_{i+1}$ implies assigning  the value of variable on lattice site $(x_i,y_i)$ to the variable on lattice site $(x_i+1,y_i)$ for all values of $(x_i,y_i)$ in one realization of the translation and  then assigning    the value of variable on lattice site $(x_i,y_i)$ to the variable on the  lattice site $(x_i,y_i+1)$, for all values of $(x_i,y_i)$ in another realization of the translation.  

Define   
\begin{eqnarray}
\sum_i S_{i+m}S_{i+n}S_{i+p}... = \sum_{k=-\infty}^{k=\infty} S_{m+k}S_{n+k}S_{p+k}..
\end{eqnarray}
  in case of one dimension.  
  
  \begin{eqnarray}
  \sum_i S_{i+m}S_{i+n}S_{i+p}.. =  \sum_{k,l=-\infty}^{k,l=\infty} S_{x_m+k,y_m + l}S_{x_n+k,y_n +l}S_{x_p+k,y_p + l}.. 
  \end{eqnarray}
  in case of two dimension and similarly for three dimensions. Note that the above terms are invariant under translation as defined above.

\section*{Proof}
The probability  for a    configuration  $\{S_i\}$ is denoted by  $f(\{S_i\})$.  The ensemble average of any quantity which is a function of $\{S_i\}$, $O(\{S_i\})$, is defined as

\begin{eqnarray}
  \langle O(\{S_i\})  \rangle  &=&  \sum_{\{S_i \}} \quad O(\{S_i\}) \quad f(\{S_i\}). \nonumber\\
\end{eqnarray}
Here $  \sum_{\{S_i \}}$ refers to summing over all possible configurations.

 Observed translational invariance of the system implies the translational invariance of the correlation functions. With this in mind,  we make the following proposition,

\vspace{5pt}
\textbf{Proposition 1:}

\emph{ Since correlation functions  are translation invariant,  the probability distribution $ f(\{S_i\})$ has to be translation  invariant.}

\vspace{5pt}
If
 \begin{eqnarray}
 \langle S_a \rangle  &=&  \sum_{\{S_i \}} \quad S_a \quad f(\{S_i\}). \nonumber\\
  \langle S_a S_b \rangle  &=&  \sum_{\{S_i \}} \quad S_a S_b \quad f(\{S_i\}). \nonumber\\
   \langle S_a S_b S_c\rangle  &=&  \sum_{\{S_i \}} \quad S_a S_b S_c\quad f(\{S_i\}). \nonumber\\
   && ...\nonumber\\
   && ....\nonumber\\
 \end{eqnarray}
 are all translation invariant, we have to have 
     that $ f(\{S_i\})$ is invariant under $S_i \rightarrow S_{i+1}$.  

\vspace{20pt}

\textbf{ Proposition 2:}

\emph{ Translation invariance coupled with discrete rotational invariance   implies full rotational invariance of the two point correlation function.}

\vspace{20pt}

Consider a two dimensional case. Let two points on the lattice be $i=(m_1,n_1)$ and $j=(m_2, n_2)$. Translation invariance   implies the two point correlation function has to have the functional form $C(i,j) = g(m_1-m_2, n_1-n_2)$. A discrete rotation is done by a matrix  

\begin{gather}
\begin{bmatrix}
a &
b\\
c &
d 
\end{bmatrix}
\end{gather}
where $a,b,c,d$ are constants. The discrete rotation symmetry which will determine the  $a,b,c,d$, would be a subgroup of the symmetry group of the lattice. Under a discrete rotation we have 
\begin{gather}
\begin{bmatrix}  m_1'-m_2' \\ n_1'-n_2' \end{bmatrix}
=
\begin{bmatrix}
a &
b\\
 c &
 d 
\end{bmatrix}
\begin{bmatrix}  m_1-m_2\\ n_1-n_2\end{bmatrix}
\end{gather}

 In order to have that $ g(m_1-m_2, n_1-n_2) = g(m_1'-m_2', n_1'-n_2')$, for \textbf{any} lattice points $i=(m_1,n_1)$ and $j=(m_2, n_2)$, we have to have that,  $ g(m_1-m_2, n_1-n_2) $ is a functional of $\sqrt{(m_1-m_2)^2+  (n_1-n_2)^2 } $, implying full rotational invariance of the two point correlation function. Since, the systems we are interested in are scale invariant,   the    correlation function should have  the functional form $C(|a-b|)= \frac{C}{|a-b|^\alpha}$. 
 
\vspace{20pt}
 \textbf{ Proposition 3: }
 
\emph{ 
  Translation invariance of  $ f(\{S_i\})$, implies  that generically $ f(\{S_i\})$  is a function of translationally invariant terms that cannot be expressed in terms of each other. Some examples of these terms  are   $ \sum_i S_i$,  $ \sum_i S_{i+n} S_{i+m} $, $\sum_i S_{i+p} S_{i+m} S_{i+n}$ etc }
\vspace{20pt}

 
 Next, because none of the $ \sum_i S_i$,  $ \sum_i S_{i+n} S_{i+m} $, $\sum_{i} S_{i+p} S_{i+m} S_{i+n}$....  etc can be broken up into a product of two terms, it implies they cannot be written in terms of each other.

 We can hence express as a Fourier transform
\begin{eqnarray}
  f(\{S_i\}) &=&   \int \Pi_i dJ_i  \quad a(J_1, J_2, J_3... )  \quad e^{i J_1  \sum_i S_i + i  \sum_{m,n }J_2^{m,n }  \sum_i S_{i+n} S_{i+m} + i\sum_{m,n,p}J_3^{m,n,p}  \sum_i S_{i+p} S_{i+m} S_{i+n} ....}\nonumber\\
  && + (*),\nonumber \\
  \label{rotational use as}
\end{eqnarray}
    
   $(*)$ in this text refers to complex conjugation of all terms to the left.  The $...$ in the exponential above   includes  all other translationally invariant terms that cannot be written in terms of each other.  Each of them multiplied by a corresponding $J_i$. In the integration we do not explicitly write the $m,n,p..$ etc labels on the $J_i$'s, but their presence is understood as we are integrating over all possible $J$'s. To regularize terms that may be ill defined because of  integrating over the $J_i$'s, we use the epsilon prescription.

\begin{eqnarray}
    f(\{S_i\}) &=&   \int \Pi_i dJ_i  \quad a(J_1, J_2, J_3... )  \quad e^{i   (1+i\epsilon\Gamma)J_1 \sum_i S_i + i  \sum_m  (1+i\epsilon \Gamma) J_2^{m,n} \sum_i S_{i+n} S_{i+m} + i\sum_{m,n} (1+i\epsilon \Gamma) J_3^{m,n,p} \sum_i S_{i+p} S_{i+m} S_{i+n} ....} \nonumber\\
    &&+ (*).\nonumber \\
  \label{eqn2}
\end{eqnarray}
    
    The operator $\Gamma$ acting on anything to its right   is defined as 
    \begin{eqnarray}
    \Gamma x &=& + x \quad x > 0 \nonumber \\
        \Gamma x &=& - x \quad x < 0, \nonumber \\
        \label{Gamma}
    \end{eqnarray}
    $ (1+i\epsilon \Gamma)$ is added to regularize the exponentials. \emph{ We take $\epsilon \rightarrow 0^+$ in the end of any calculation}. This is inspired by the method of  regularizing a path integral by taking the time co-ordinate to have  a slight imaginary component. 

  Now,
 
\begin{eqnarray}
 C(|a-b|) &=&   \langle S_a S_b \rangle - \langle S_a \rangle\langle  S_b \rangle \nonumber \\
 &=&   \int \sum_{\{S_i\}} \Pi_j dJ_j  \quad a(J_1, J_2, ... )S_a S_b   \nonumber \\
 &\times& \quad e^{i   (1+i\epsilon\Gamma) J_1\sum_i S_i + i  \sum_m  (1+i\epsilon\Gamma) J_2^{m,n} \sum_i S_{i+n} S_{i+m} + i\sum_{m,n}  (1+i\epsilon\Gamma)J_3^{m,n}\sum_i S_{i+p} S_{i+m}    S_{i+n} ....} + (*) - \langle S_a \rangle\langle  S_b \rangle\nonumber\\
  &=&  \langle S_a S_b \rangle - \langle S_a \rangle\langle  S_b \rangle \nonumber \\
  &=&   \int   \Pi_j dJ_j  \quad a( J_1, J_2, ... ) \quad Z( J_1, J_2,  ... )\langle S_a S_b \rangle_{ J_1, J_2,  ...}   +(*) - \langle S_a \rangle\langle  S_b \rangle\nonumber\\
 &=&    \underbrace{ \int  \Pi_j dJ_j a( J_1, J_2,... ) Z(J_1, J_2,...)C(a,b)_{J_1,J_2,J_3, ..} }_{ function\; of \; position's \; a \; and \; b}\nonumber \\
 &+&    \int \Pi_j dJ_j  Z(J_1, J_2,...)a(J_1, J_2, ...  )\langle S_a \rangle_{J_1,J_2,J_3..}\langle  S_b \rangle _{J_1,J_2,J_3..} + (*)  \nonumber\\
  &-& \langle S_a \rangle\langle  S_b \rangle  ,\nonumber\\
  \label{2point}
\end{eqnarray}
 Where, we  have defined

\begin{eqnarray}
Z(J_1, J_2...) &=&   \sum_{\{S_i\}}   e^{i   (1+i\epsilon \Gamma) J_1\sum_i S_i + i  \sum_m   (1+i\epsilon\Gamma)J_2^{m,n} \sum_i S_{i+n} S_{i+m} + i\sum_{m,n ,p} (1+i\epsilon\Gamma)J_3^{m,n,p} \sum_{i} S_{i+p} S_{i+m} S_{i+n} ....}  \nonumber \\
\langle   O( {\{S_i\}} ) \rangle_{J_1,J_2,J_3..}  &=&    \frac{ \sum_{\{S_i\}} O( {\{S_i\}} )   e^{i   (1+i\epsilon\Gamma) J_1\sum_i S_i + i  \sum_{m,n}   (1+i\epsilon\Gamma) J_2^{m,n}\sum_i S_{i+n} S_{i+m} + i\sum_{m,n,p}  (1+i\epsilon\Gamma) J_3^{m,n,p}\sum_i S_{i+p} S_{i+m} S_{i+n} .... ....}  }{ Z(J_1, J_2...) }\nonumber \\
 C(a,b)_{J_1,J_2,J_3..}  &=& \langle S_a S_b \rangle_{J_1,J_2,J_3..}  - \langle S_a  \rangle_{J_1,J_2,J_3..}   \langle   S_b\rangle _{J_1,J_2,J_3..},
\end{eqnarray}
and $ O( {\{S_i\}} ) $ is any functional of the  $S_i$ configuration on the lattice. 
	Since $ \langle S_a \rangle$ ,  $\langle S_a \rangle_{J_1,J_2,J_3..}$, $\langle  S_b \rangle _{J_1,J_2,J_3..}$ are independent of  lattice sites $a,b$  we have
	\begin{eqnarray}
	\int \Pi_j dJ_j a(J_1, J_2... )Z(J_1, J_2...)\langle S_a \rangle_{J_1,J_2,J_3..}\langle  S_b \rangle _{J_1,J_2,J_3..} + (*)
- \langle S_a \rangle\langle  S_b \rangle = Ind(a,b) ,\nonumber\\
 \label{eq1}
	\end{eqnarray}
	where, $Ind(a,b)$   does not depend on $ a$ or $b $.

Since $C(|a-b|)$  is rotationally invariant, it is not possible for    $ C(a,b)_{J_1,J_2,J_3..} $ to be not  rotationally invariant for all possible combinations of $a$ and $b$.  The power of this statement cannot be understated. Rotational invariance of $C(|a-b|)$, sets only those $a( J_1, J_2,... )$'s to be non zero for which $ C(a,b)_{J_1,J_2,J_3..} $  is rotationally invariant. Because of rotational invariance, we can   write $C(a,b)_{J_1,J_2,J_3..} = C(|a-b|)_{J_1,J_2,J_3..}$. 

\vspace{20pt}	
	\textbf{ Proposition 4:}
		\emph{ If $|a-b|$   is order's of magnitude greater than  the lattice spacing
		\begin{eqnarray}
	 C(|a-b|)_{J_1,J_2,J_3..} &\sim& \frac{e^{ - \lambda_ {J_1,J_2,J_3..}|a-b| }}{|a-b|^{\alpha_{J_1,J_2,J_3..} }}.
		\end{eqnarray}	
	}
	\vspace{20pt}	
	with $ \lambda_ {J_1,J_2,J_3..} \ge 0$. 
	
	Consider
	\begin{eqnarray}
	1 &=& N \int \Pi_i D\psi_i e^{-iJ_1 \sum_i \psi_i -i \sum_{m} J_2^m  \sum_i \psi_i \psi_{i+m} - i \sum_{m,n} J_3^{m,n}  \sum_i \psi_i \psi_{i+m} \psi_{i+n}}... \nonumber \\
	\end{eqnarray}
	$\psi \in [-\infty, \infty]$ during integration. $N $ is an normalization constant. Finiteness of the integral is obtained by an epsilon prescription as was done in Eq.\ref{eqn2} and Eq.\ref{Gamma}. The $J$'s above have to be read as being multiplied by $ (1+i\epsilon\Gamma)$ (we do not write this multiplication explicity to prevent the equations below from appearing too messy).  Redefining $\psi_i \rightarrow \psi_i + S_i$, we get

	\begin{eqnarray}
	 1&=&N  \int \Pi_i D\psi_i e^{-iJ_1 \sum_i \psi_i -iJ_1 \sum_i S_i-i\sum_m J_2^m  \sum_i \psi_i \psi_{i+m} - i \sum_{m,n} J_3^{m,n}  \sum_i \psi_i \psi_{i+m} \psi_{i+n}} \nonumber \\
	&\times&    e^{  -2i \sum_m J_2^m   \sum_i S_i \psi_{i+m}  -i\sum_m J_2^m  \sum_i S_i S_{i+m} } \nonumber \\
	&\times&  e^{  -3 i \sum_{m,n} J_3^{m,n}  \sum_i S_i \psi_{i+m} \psi_{i+n} -3 i \sum_{m,n} J_3^{m,n}  \sum_i S_i S_{i+m} \psi_{i+n}- i \sum_{m,n} J_3^{m,n}  \sum_i S_i S_{i+m} S_{i+n}...} \nonumber \\
	\end{eqnarray}
	or,
		\begin{eqnarray}
	&&e^{i J_1  \sum_i S_i + i  \sum_m J_2^m  \sum_i S_i S_{i+m} + i\sum_{m,n}J_3^{m,n}  \sum_i S_i S_{i+m} S_{i+n} ....} \nonumber\\
	 &=& N  \int \Pi_i D\psi_i e^{-iJ_1 \sum_i \psi_i  -i\sum_m J_2^m  \sum_i \psi_i \psi_{i+m} - i \sum_{m,n} J_3^{m,n}  \sum_i \psi_i \psi_{i+m} \psi_{i+n}} \nonumber \\
	&\times&    e^{  -2i \sum_m J_2^m   \sum_i S_i \psi_{i+m}   } \nonumber \\
	&\times&  e^{  - 3i \sum_{m,n} J_3^{m,n}  \sum_i S_i \psi_{i+m} \psi_{i+n} - 3i \sum_{m,n} J_3^{m,n}  \sum_i S_i S_{i+m} \psi_{i+n} ...}. \nonumber \\
	\end{eqnarray}
	
	So, 
	\begin{eqnarray}
	Z&= &\Pi_{\{S_i\}} e^{i J_1  \sum_i S_i + i  \sum_m J_2^m  \sum_i S_i S_{i+m} + i\sum_{m,n}J_3^{m,n}  \sum_i S_i S_{i+m} S_{i+n} ....} \nonumber\\
	 &=& N \Pi_{\{S_i\}}  \int \Pi_i D\psi_i e^{-iJ_1 \sum_i \psi_i  -i\sum_m J_2^m  \sum_i \psi_i \psi_{i+m} - i \sum_{m,n} J_3^{m,n}  \sum_i \psi_i \psi_{i+m} \psi_{i+n}} \nonumber \\
	&\times&    e^{  -2i \sum_m J_2^m   \sum_i S_i \psi_{i+m}   } \nonumber \\
	&\times&  e^{  - 3i \sum_{m,n} J_3^{m,n}  \sum_i S_i \psi_{i+m} \psi_{i+n} - 3 i \sum_{m,n} J_3^{m,n}  \sum_i S_i S_{i+m} \psi_{i+n} ...}. \nonumber \\
	\end{eqnarray}
The RHS in the second equation after summing over all $S_i$'s can be written as $e^{-i S(..\psi_{i}, \psi_{i+1},...)}$, where $S(..\psi_{i}, \psi_{i+1},...)$ is a function of all values of $\psi_{i}$'s. Because every term in every exponent in the integral above is translation invariant $ S(..\psi_{i}, \psi_{i+1},...)$ is translation invariant, in the continuum limit (corresponding to considering distances much  larger than the lattice spacing, where the lattice appears as a continuum)  we have 
	\begin{eqnarray}
	S &=& \int d^d x [c_1 \bigtriangledown \psi \cdot \bigtriangledown \psi + c_2 \psi^2 +\sum_{m,n.p} c_{mnp} \psi^m  (\bigtriangledown \psi \cdot \bigtriangledown \psi )^n  \bigtriangledown^{2p} \psi],
	\label{Lagrangian}
	\end{eqnarray}
  The $c$'s are functions of the $J$'s.  (Eq. 15 to Eq. 19 are inspired from  Eq. 5.5 to Eq 5.6 from \cite{atland}, where the transformation are done for partition functions).
   Rotational invariance of the expression above is necessary in  ensure  that  $ C(|a-b|)_{J_1,J_2,J_3..} $   is rotationally invariant.   The above form for $S$ is set by including      all possible rotationally invariant terms, which should be allowed in the continuum limit.   It is well known that for a system described by a field theory, the correlation function at distances $r$  orders of magnitude larger than the lattice spacing  scales as  
	
	\begin{eqnarray}
	\sim \frac{e^{-\lambda_{ J_1,J_2,J_3..}  r} }{r^{\alpha_{ J_1,J_2,J_3..} }}.
	\end{eqnarray} 

The reason behind this is   for large distances (infrared limit), higher derivative terms and higher powers of $\psi$       contribute minisculely in any effective field theory \cite{eft}. 	Hence, the correlation function is just one that is obtained from an action $ \int d^d x [c_1 \bigtriangledown \psi \cdot \bigtriangledown \psi + c_2 \psi^2+...] $, where the $...$ are the relevant and marginal operators. The correlation function   has a form as above. 

	This furnishes   proof of  $C(|a-b|)_{J_1,J_2,J_3..} \sim \frac{e^{ - \lambda_ {J_1,J_2,J_3..}|a-b| }}{|a-b|^{\alpha_{J_1,J_2,J_3..} }} $ at large $|a-b|$.

 {\textbf{Proposition 5}: $ Ind(a,b) = 0$}

Proof :  

Since our system is scale invariant $ C(|a-b|) = \frac{C}{|a-b|^\alpha}$, where $C$ is a constant  and $|a-b|$ is quite large  
\begin{eqnarray}
\frac{C}{|a-b|^\alpha}
&=&  \int  \Pi_j dJ_j  a(J_1, J_2... )Z(J_1, J_2...)\frac{e^{ - \lambda_ {J_1,J_2,J_3..}|a-b| }}{|a-b|^{\alpha_{J_1,J_2,J_3..} }} + (*)+Ind(a,b).  \nonumber\\
\label{eqn1}
\end{eqnarray}	
In the above we have absorbed the constant of proportionality in $	 C(|a-b|)_{J_1,J_2,J_3..} \sim  \frac{e^{ - \lambda_ {J_1,J_2,J_3..}|a-b| }}{|a-b|^{\alpha_{J_1,J_2,J_3..} }}$, into $ a(J_1, J_2... )$. Now, take $|a-b| \rightarrow \infty$. Then, since first term on RHS is zero and LHS is zero, it implies $Ind(a,b)  = 0$
since $Ind(a,b) $ does not depend on either $a$ or $b$.

\vspace{30pt}
\textbf{Proposition 6:} 

\emph{$\alpha_{J_1,J_2,J_3..} = \alpha $,   if $a( J_1, J_2... ) \neq 0$. }
\vspace{20pt}

Proof:
   Differentiate both sides of Eq.\ref{eqn1} with respect to $|a-b|$. This would give
\begin{eqnarray}
 \frac{-\alpha C}{|a-b|^{\alpha+1}}
 &=&  \int  \Pi_j dJ_j  a( J_1, J_2... )Z(J_1, J_2...)\frac{-\alpha_{J_1,J_2,J_3..} e^{ -  \lambda_ {J_1,J_2,J_3..}|a-b| }}{|a-b|^{\alpha_{J_1,J_2,J_3..} +1}}   \nonumber\\
 &-&  \int  \Pi_j dJ_j  a( J_1,J_2... )Z(J_1, J_2...)\frac{\lambda_ {J_1,J_2,J_3..} e^{ -  \lambda_ {J_1,J_2,J_3..}|a-b| }}{|a-b|^{\alpha_{J_1,J_2,J_3..} }}  + (*) \nonumber\\
 &=&  -\int  \Pi_j dJ_j  a( J_1,J_2... )Z(J_1, J_2...)\frac{\alpha e^{ - \lambda_ {J_1,J_2,J_3..}|a-b| }}{|a-b|^{\alpha_{J_1,J_2,J_3..}+1 }}   + (*). \nonumber\\
 \label{below}
\end{eqnarray}

To see   the last step, equalize the RHS of the first and second line of the above equation to the RHS of Eq.\ref{eqn1} after multiplying on both sides by $\frac{-\alpha }{|a-b|}$.
 Consistency for all possible  values of $|a-b|$ in the last equation (which are still large enough), requires $\int  \Pi_j dJ_j  a(...J_i, J_{i+1}... )Z(J_1, J_2...)\frac{\lambda_ {J_1,J_2,J_3..} e^{ -  \lambda_ {J_1,J_2,J_3..}|a-b| }}{|a-b|^{\alpha_{J_1,J_2,J_3..} }} + (*)= 0$, as it dominates over $  \Pi_j dJ_j  a( J_1,J_2... )Z(J_1, J_2...)\frac{-\alpha_{J_1,J_2,J_3..} e^{ -  \lambda_ {J_1,J_2,J_3..}|a-b| }}{|a-b|^{\alpha_{J_1,J_2,J_3..} +1}}$.

 Hence we get
\begin{eqnarray}
 \frac{-\alpha C}{|a-b|^{\alpha+1}}
 &=&  \int  \Pi_j dJ_j  a( J_1,J_2... )Z(J_1, J_2...)\frac{-\alpha_{J_1,J_2,J_3..} e^{ -  \lambda_ {J_1,J_2,J_3..}|a-b| }}{|a-b|^{\alpha_{J_1,J_2,J_3..} +1}} + (*) \nonumber\\
 &=&  -\int  \Pi_j dJ_j  a( J_1,J_2... )Z(J_1, J_2...)\frac{\alpha e^{ - \lambda_ {J_1,J_2,J_3..}|a-b| }}{|a-b|^{\alpha_{J_1,J_2,J_3..}+1 }}  + (*).\nonumber\\
\end{eqnarray}	
Differentiating again with respect to $a$ similarly gives
\begin{eqnarray}
\frac{ \alpha(\alpha+1)C}{|a-b|^{\alpha+2}}
 &=&  \int  \Pi_j dJ_j  a( J_1,J_2... )Z(J_1, J_2...)\frac{ \alpha_{J_1,J_2,J_3..} (  \alpha_{J_1,J_2,J_3..} +1 )e^{ -  \lambda_ {J_1,J_2,J_3..}|a-b| }}{|a-b|^{\alpha_{J_1,J_2,J_3..} +2}} \nonumber\\
 &+&  \int  \Pi_j dJ_j  a( J_1,J_2... )Z(J_1, J_2...)\frac{\lambda_ {J_1,J_2,J_3..} \alpha_{J_1,J_2,J_3..}  e^{ -  \lambda_ {J_1,J_2,J_3..}|a-b| }}{|a-b|^{\alpha_{J_1,J_2,J_3..}+1}}+ (*) \nonumber\\
  &=&  \int  \Pi_j dJ_j  a( J_1,J_2... )Z(J_1, J_2...)\frac{ \alpha_{J_1,J_2,J_3..} (  \alpha_{J_1,J_2,J_3..} +1 )e^{ -  \lambda_ {J_1,J_2,J_3..}|a-b| }}{|a-b|^{\alpha_{J_1,J_2,J_3..} +2}}, \nonumber\\
\end{eqnarray}	

where last line is because   of multiplying Eq.\ref{eqn1} by $ \alpha(\alpha+1) $. We hence get by  arguments similar to ones below Eq.\ref{below}.

\begin{eqnarray}
&& \int  \Pi_j dJ_j  a( J_1,J_2... )Z(J_1, J_2...)\frac{ \alpha_{J_1,J_2,J_3..} (  \alpha_{J_1,J_2,J_3..} +1 )e^{ -  \lambda_ {J_1,J_2,J_3..}|a-b| }}{|a-b|^{\alpha_{J_1,J_2,J_3..} +2}} \nonumber\\
&=&   \int  \Pi_j dJ_j  a( J_1,J_2... )Z(J_1, J_2...)\frac{\alpha (\alpha+1) e^{ - \lambda_ {J_1,J_2,J_3..}|a-b| }}{|a-b|^{\alpha_{J_1,J_2,J_3..}+2 }}   + (*).\nonumber\\
\end{eqnarray}	
 One can keep taking derivatives to get
 \begin{eqnarray}
\frac{ P(\alpha, n) C}{|a-b|^{\alpha+n}}
 &=&  \int  \Pi_j dJ_j  a( J_1,J_2... )Z(J_1, J_2...)\frac{ P(\alpha_{J_1,J_2,J_3..},n) e^{ -  \lambda_ {J_1,J_2,J_3..}|a-b| }}{|a-b|^{\alpha_{J_1,J_2,J_3..} +n}} \nonumber\\
 &+&  \int  \Pi_j dJ_j  a(...J_i, J_{i+1}... )Z(J_1, J_2...)\frac{\lambda_ {J_1,J_2,J_3..} P(\alpha, n-1)e^{ -  \lambda_ {J_1,J_2,J_3..}|a-b| }}{|a-b|^{\alpha_{J_1,J_2,J_3..}+n-1}} + (*) \nonumber\\
  &=&  \int  \Pi_j dJ_j  a( J_1,J_2... )Z(J_1, J_2...)\frac{ P(\alpha_{J_1,J_2,J_3..},n) e^{ -  \lambda_ {J_1,J_2,J_3..}|a-b| }}{|a-b|^{\alpha_{J_1,J_2,J_3..} +n}} + (*)\nonumber\\
\end{eqnarray}

and hence

 \begin{eqnarray}
&& \int  \Pi_j dJ_j  a( J_1,J_2... )Z(J_1, J_2...)\frac{ P(\alpha_{J_1,J_2,J_3..},n) e^{ -  \lambda_ {J_1,J_2,J_3..}|a-b| }}{|a-b|^{\alpha_{J_1,J_2,J_3..} +n}} + (*)\nonumber\\
&=&   \int  \Pi_j dJ_j  a( J_1,J_2... )Z(J_1, J_2...)\frac{ P(\alpha, n)  e^{ - \lambda_ {J_1,J_2,J_3..}|a-b| }}{|a-b|^{\alpha_{J_1,J_2,J_3..}+n }}   + (*),\nonumber\\
\end{eqnarray}
where $P(x, n) = x (x + 1)...(x + n-1)$ 	
 for $n>0$ and $P(x, n=0) = 1$ and $P(x, n=-1) = 0$. Only way the above equation is possible for all large values of $|a-b|$ and all $n > 0 $ is if $ \alpha_{J_1,J_2,J_3..}   = \alpha$  for $ a(...J_i, J_{i+1}... ) \neq 0 $ .

\vspace{20pt}
\textbf{Proposition 7: } 

Atleast one $\lambda_ {J_1,J_2,J_3..} = 0$, for $ a(...J_i, J_{i+1}... ) \neq 0 $.

\vspace{20pt}

Proof:  From Proposition 6, we get that  
\begin{eqnarray}
  \int  \Pi_j dJ_j  a(J_1, J_2... )Z(J_1, J_2...) e^{ - \lambda_ {J_1,J_2,J_3..}|a-b| } + (*) = C  \nonumber\\
\end{eqnarray}	
where $C$ is constant for all values of $|a-b| $. If $|a-b| \rightarrow \infty$, this constant is equal to zero in case all  of the $\lambda_ {J_1,J_2,J_3..} \neq 0$ as all  $\lambda_ {J_1,J_2,J_3..} > 0$. Hence atleast one $\lambda_ {J_1,J_2,J_3..} = 0$, for $ a(...J_i, J_{i+1}... ) \neq 0 $, if we have to have a non zero value for $C$.  Having a particular  $\lambda_ {J_1,J_2,J_3..} = 0$,  implies were are really talking about the critical point of a field theory with an action of the form Eq.\ref{Lagrangian}. Hence, the critical exponent $\alpha$ is the scaling exponent for such a field theory.  At criticality such models fall into universality classes, with models in  each universality class having the same critical exponent.   This completes our proof that  the scaling critical exponent in the lattice model described by an arbitrary probability distribution will match the scaling exponent of  a  statistical mechanical model in a universality class. Implicit in this statement is   the observation that the scaling exponent of the correlation function in   a quantum field theory is   same as in a corresponding statistical field theory at the fixed point. 

 \begin{figure}
	\includegraphics[scale = .25]{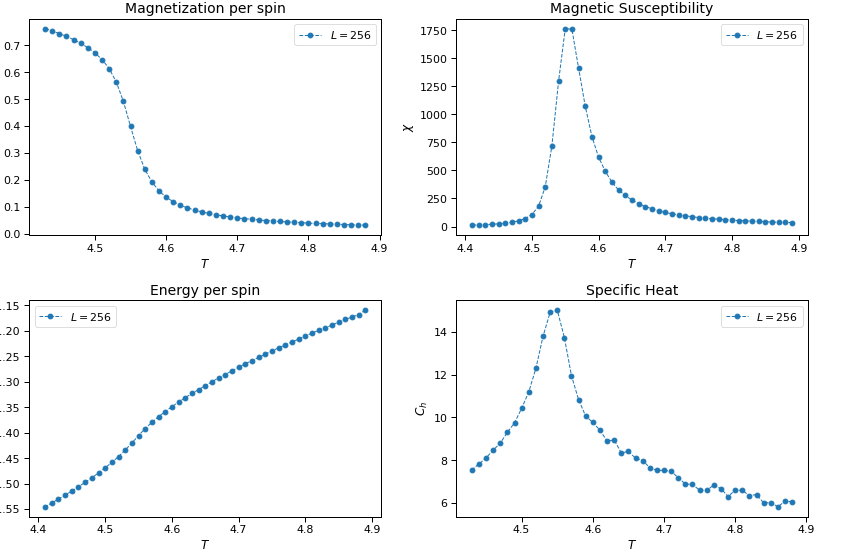}
	\caption{ }
	\label{fig1}
\end{figure}

\begin{figure}
	\includegraphics[scale = .25]{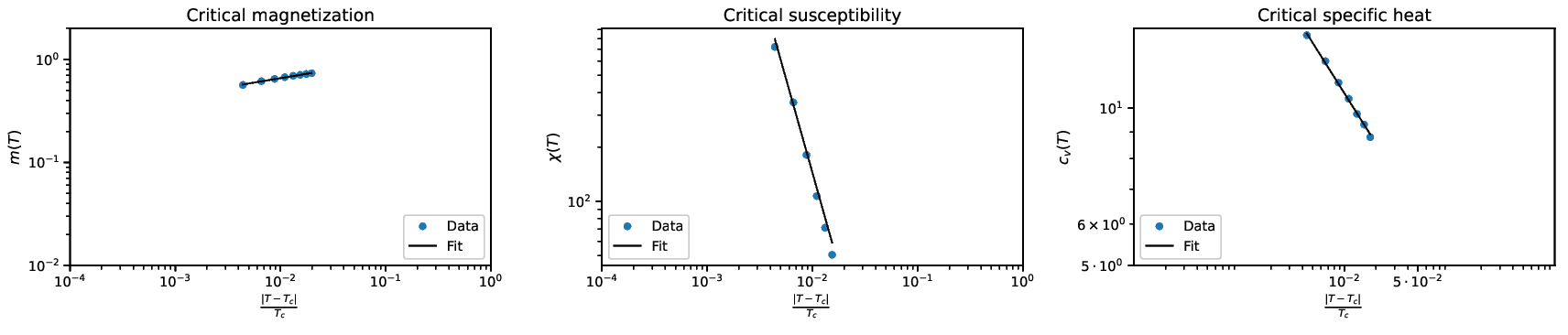}
	\caption{ }
	\label{fig2}
\end{figure}
\section*{Discussion}
{ One can consider a Monte Carlo simulation to test the hypothesis in the paper. The reason one does a Monte Carlo simulation in these problems is because the phase space of all possiblities configurations possible to the system is extremely large and the best bet is hence to consider Monte Carlo simulations which are akin to time averages. The hope is that in a time average of a substantially smaller number of steps than the total number of configurations, one could have time averages mimicking the ensemble averages.   We consider the following probability distribution
\begin{equation}
P(\{S_i\}) = [e^{-\sum_{\langle I,J \rangle} \frac{S_I S_I}{T}} +e^{-\sum_{\langle I,J \rangle} \frac{2 S_I S_I}{T}}  ]
\end{equation}
  on a square lattice with  $256 \times 256$ points and $S_I \in [0,1]$ at each lattice point takes integer values. At each step a site is chosen at random and $S_I$   flipped if probability of the new configuration is larger than the previous one, or flipped if    a random number chosen  $\in [0,1]$ is larger than the ratio of the  probability of the pre-flip configuration to the post flip configuration.  We changed a freely available code \cite{github} which was made to simulate a 2D Ising model by modifying it to our needs. Calling $S_I$ the magnetization order parameter just like the Ising model, we get the   graphs   shown in Fig.\ref{fig1} and Fig.\ref{fig2}. Taking the critical $T_c = 4.55$, we evaluate $\beta = -.17$, $\gamma = 1.9$ and $\alpha = .32$ from slope of the curves in Fig. \ref{fig2}. Using the formula $\frac{2-\eta_\alpha}{d}  = \frac{\gamma}{2-\alpha}$ we find the scaling  exponent to be  $.25$ as is of an Ising model, confirming our hypothesis that the scaling exponent is universal irrespective of the probability distribution. However this is to be expected in the simulation. The reason being that the probability has a form $e^{-a E} + e^{-2 a E}$ where $a>0$ is a constant and $E$ is the "energy" of the configuration. Since in the thermodynamic limit $E$ goes as the system size which is very large, the $e^{-2 a E}$ in evaluation of any averages  (corresponding to $E<0$ configurations) dominates the $e^{-a E}$ term. If one instead considers a probability distribution like 
\begin{equation}
	P(\{S_i\}) = \cos[ {\sum_{\langle I,J \rangle} \frac{S_I S_I}{T}}   ]
\end{equation}
which would not have issues of such a single exponential term dominating, one ends up with an issue that than the ratio of the  probability of the pre-flip configuration to the post flip configuration is always close to one for large system size, making the Monte Carlo method of time averaging not being able to reproduce a ensemble average in finite tractable computational steps. We note that in order for Monte Carlo methods to work we need that the ratio of probability of the configuration before a flip $P(E_b)$ to the probability of a configuration   after the flip $P(E_a)$ to not be close to one. Here $E_a, E_b$ correspond to a quantity that is a functional of the configuration akin to energy in equilibrium statistical mechanics.  $E_a, E_b$  go as the size of the configuration $N$. Hence $|\frac{E_b -E_a}{E_a}| \sim \mathcal{O}(1/N)$. We need $\frac{P(E_b)}{P(E_a)}   $ to be a function of $E_b - E_a$ in order to get  $\frac{P(E_b)}{P(E_a)}  $ to a value which is not close to one.  This necessitates a exponential dependence on $E_a$ or $E_b$ of the probability distribution. It is here where our proof above becomes extremely important, because even in case where a Monte Carlo simulation may not be computationally tractable or even cases where the probability distribution producing the data may not even be known,   the  scaling exponent in cases where the   the data shows scale invariance of the correlation function, can be confirmed to lie  in a equilibrium statistical model. 
}
 
We   note that what is said above is only confirmed for systems defined on a lattice, where the variable defining the state of the system at a lattice site   takes the same possible values on every lattice site. These values  are also finite.  The result that  the way the correlations   decay in systems having long range order described by any arbitrary probability distributions,  are similar to statistical models on a lattice at criticality  is a new addition to the understanding of universality classes in physics. Critical nature has been hypothesized in non statistical mechanical systems going from retinal neurons \cite{bialek}, to natural images \cite{images}, which could be represented as systems on a lattice. Our result above says that irrespective of the actual probability distributions describing the statistics of  these systems, the scaling exponents if measured would fall into universality classes of statistical mechanical models.  The result in the paper  could also aid in the working out analytically the critical exponents  of   of statistical mechanical models at criticality, by considering  appropriate alternative probability distributions, instead of the Boltzmann distribution in which the calculation is tedious.

 \section*{Additional Information}
 There are no competing interests.
 
 \section*{ Data Availability}
 Data sharing not applicable – no new data generated.
 \section*{Acknowledgements}
 We would like to thank Prof Ranjan Mukhopadhyay for useful comments on the manuscript.

\section*{Compliance with Ethical Standards}
\noindent Funding:       None.\\
Conflict of Interest: None\\
Ethical Conduct: Work was done keeping in mind scientific ethics.

\end{document}